\documentclass{jetpl}
\input{psfig}
\twocolumn
\lat
\newcommand{\ket}[1]{\left|#1\right\rangle}
\title{Quantum Zeno effect in the Cooper-pair transport\\
through a double-island Josephson system}
\rtitle{Quantum Zeno effect in the Cooper-pair transport\ldots}
\sodtitle{Quantum Zeno effect in the Cooper-pair transport
through a double-island Josephson system}
\author{Alexander~Shnirman$^+$ and Yuriy~Makhlin$^{+,*}$}
\rauthor{A.~Shnirman, Yu.~Makhlin}
\sodauthor{Shnirman~A., Makhlin~Yu.}
\address{$^+$Institut f\"ur Theoretische Festk\"orperphysik,
Universit\"at Karlsruhe, D-76128 Karlsruhe, Germany\\
$^*$L.D.~Landau Institute for Theoretical Physics,
Kosygin st. 2, 117940 Moscow, Russia}

\renewcommand\dates[2]{\relax}
\abstract{Motivated by recent experiments, we analyze transport of Cooper
pairs through a double-island Josephson qubit. At low bias in a
certain range of gate voltages coherent superpositions of charge
states play a crucial role. Analysis of the evolution of the
density matrix allows us to cover a wide range of parameters,
incl. situations with degenerate levels, when dissipation strongly
affects the coherent eigenstates. At high noise levels the
so-called Zeno effect can be observed, which slows down the
transport. Our analysis explains certain features of the $I$--$V$
curves, in particular the visibility and shape of resonant peaks
and lines.}

\begin{document}
\maketitle

Among various proposals for realization of qubits, solid-state devices appear
particularly promising since they can be easily scaled up to large qubit
registers and integrated in electronic circuits~\cite{Our_RMP}. Recent 
experiments have
demonstrated quantum coherent oscillations in Josephson-junction devices.
However, in such devices due to the host of microscopic modes decoherence
processes are more difficult to control, and understanding of the decoherence
mechanisms requires further analysis. Further, improvements of the quantum
measurement procedure are needed to allow monitoring the qubit's state with
little influence on the qubit's dynamics before the read-out. Here we analyze
recent experiments~\cite{Bibow_Lafarge_Levy_PRL,Bibow_Thesis}, in which the
dissipative dynamics of a Josephson charge qubit were probed by Cooper-pair
transport. This experiment provides data for understanding of the dissipation in
typical superconducting charge devices, and its analysis is similar to that for
the quantum charge detectors.

We focus on the analysis of Josephson circuits in the charge limit, in
which the typical electrostatic energy needed to charge a superconducting island
($\sim (2e)^2/2C_\Sigma$, where $C_\Sigma$ is the total capacitance) is  higher
than the Josephson energy which controls the charge tunneling. If the system is
biased close to a point, where two charge states with lowest
energies are degenerate, at low temperatures and operation frequencies one can
neglect the higher charge states, and the system reduces to two levels (qubit).
The matrix element between these levels is controlled by the Josephson 
tunneling.
In the simplest design, a Cooper-pair box~\cite{Bouchiat}, the quantum state of
this qubit can be manipulated by voltage and current
pulses~\cite{SSH_and_Our_Nature}. The measurement of the quantum state can be
performed, for example, by coupling the qubit to a single-electron transistor
(SET) and monitoring its current~\cite{Our_RMP}. Here we study a circuit, which
can be described as a charge qubit inside a SET. Transport in
this device probes typical time scales of the qubit dynamics, and its analysis 
may show new possibilities to perform the read-out.
Our results explain experimentally observed features of transport (the 
visibility and shape of resonant lines and peaks) and predict new
specific behavior in a low-bias regime, in which coherent
properties of the double-island qubit are probed.

\begin{figure}
\centerline{\hbox{\psfig{figure=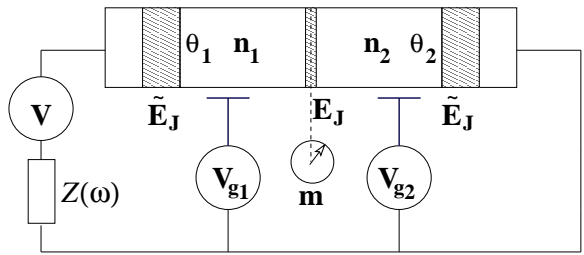,width=0.9\columnwidth}}}
\caption[]{\label{F:System}FIG.\ref{F:System}.
The double-island system}
\end{figure}

\noindent{\bf The circuit and its description.}
Following the experimental work~\cite{Bibow_Lafarge_Levy_PRL,Bibow_Thesis} we
study the system shown in Fig.~\ref{F:System}.
It consists of a Josephson junction, with a relatively strong coupling $E_{\rm 
J}$, connected to further superconducting leads via weaker junctions, with 
$\tilde E_{\rm J} \ll E_{\rm J}$. The transport is controlled by a bias $V$ 
between the 
external leads and gate voltages $V_{\rm g1}$, $V_{\rm g2}$, with gate
capacitances much lower than those of the junctions, $C_{\rm g} \ll C_{\rm J}$.
The transport of Cooper pairs through a similar system with a single island
between the leads (a superconducting SET) was studied, for instance, in
Refs.~\cite{Averin_Aleshkin,Maassen_Schoen_Geerligs_PRL_91,
Maassen_Odintsov_Bobbert_Schoen_ZPhysB_91, Siewert_Schoen_PRB_96}.  Transport
at a finite bias implies dissipation which can be provided by various
mechanisms. Here we focus on low voltages and
temperatures, at which the contribution of quasiparticles is 
negligible~\cite{Bibow_Thesis}.  We study the influence of the
electromagnetic environment, i.e., effective impedances in the circuit.  Since 
$C_{\rm g} \ll C_{\rm J}$ the impedance of the transport voltage
circuit is expected to dominate dissipation (see Fig.~\ref{F:System}).

In Refs.~\cite{Maassen_Schoen_Geerligs_PRL_91,
Maassen_Odintsov_Bobbert_Schoen_ZPhysB_91, Siewert_Schoen_PRB_96} the analysis
was limited to the evolution of occupations of the eigenstates (the diagonal
entries of the density matrix in the eigenbasis of the non-dissipative
Hamiltonian).  The dynamics were described in terms of incoherent transitions
between these states.  This approach is sufficient as long as fluctuations
provide only a weak perturbation (the incoherent rates are lower than the
coherent level splittings).  However, in a system with almost degenerate
eigenstates this approach may fail, since the system crosses over to the
so-called Zeno regime~\cite{Harris_Stodolsky_Zeno}.  To illustrate this concept
we consider a situation relevant for the analysis below:  often the charge
transport may be described as a chain of transitions between various charge
configurations.  Under certain conditions one link in this chain is a pair of
degenerate charge states, with the coupling $\delta$ between them, coupled by
incoherent transitions, with rate $\sim\Gamma$, to further states.
\label{Zeno-intro}
As long as $\delta\gg\Gamma$ transport within the pair is fast, and the current
magnitude is set by $\Gamma$.  However, if the coupling $\delta$ becomes weaker
than $\Gamma$, the dynamics change dramatically:  frequent `observation' (fast
dephasing) by the transitions destroys the coherence and slows down the
evolution; the system is blocked for a long time in one of the charge states in
the pair, with the typical transition rate $\sim\delta^2/\Gamma$, which now sets
the current magnitude.  The density matrix of the two-state system quickly
becomes diagonal in the charge basis, while in the eigenbasis diagonal and
off-diagonal entries of the density matrix are strongly coupled
(cf.~Ref.~\cite{Our_RMP}).  In order to describe the behavior of the system in
both limits, we analyze the system using the master equation for the evolution
of all entries of the density matrix.

We describe the state of the system by the charges $en_1$, $en_2$ of the central 
islands and introduce the charge $em$ transferred across the system of three 
junctions (see below for a precise definition). In the Hamiltonian,
\begin{equation}
H = H_{\rm C} + H_{\rm J} + H_{\rm diss} \,,
\label{eq:ham}
\end{equation}
the charging part is given by
\begin{equation}
\label{eq:Charging_Part}
H_{\rm C}=
\frac{( en_- + C_{\rm g}V_{\rm g-} )^2}{4(3C_{\rm J} + C_{\rm g})} +
\frac{( en_+ + C_{\rm g}V_{\rm g+} )^2}{4(C_{\rm J} + C_{\rm g})}
- Q_{\rm int} V ,
\end{equation}
where $n_\pm\equiv n_1\pm n_2$, $V_{\rm g\pm} \equiv V_{{\rm g}1} \pm V_{{\rm 
g}2}$ and
\begin{equation}
Q_{\rm int} \equiv
\frac{en_-(2C_{\rm J} + C_{\rm g})}{2(3C_{\rm J} + C_{\rm g})}
+ \frac{en_+C_{\rm g}}{2(C_{\rm J} + C_{\rm g})} + em
\end{equation}
is the charge operator that couples to the voltage source.
The Josephson part of the Hamiltonian is
\begin{equation}
H_{\rm J}=-\tilde E_{\rm J} (\cos\theta_1 + \cos\theta_2)
- E_{\rm J}\cos(\theta_2\! -\! \theta_1 \!+\! \Psi_m)
\,,
\label{eq:Josephson Energy}
\end{equation}
where $\theta_1$, $\theta_2$ are the phase drops across the left and the right
junctions, respectively, and $\exp(i\Psi_m):\ket{m} \mapsto \ket{m+2}$ is the 
counting ladder operator. One can see from
Eq.~(\ref{eq:Josephson Energy}) that tunneling of a Cooper pair
across the central junction changes $m$ by 2.

Finally, the dissipative part of the Hamiltonian
reads (cf.~\cite{Ingold_Nazarov_Book,Weiss_Book}):
\begin{equation}
\label{eq:Dissipative_Hamiltonian}
H_{\rm diss} = \frac{(Q_{\rm int} - q)^2}{2C_{\rm int}} + 
\sum_\alpha \left[ \frac{q_\alpha^2}{2C_\alpha}
+ \frac{\hbar^2}{e^2} \frac{(\phi_\alpha - \phi)^2}{2L_\alpha} \right]\,.
\end{equation}
Here $\phi$ is the phase drop across the impedance $Z(\omega)$ and $C_{\rm int}$ 
is the capacitance between its leads.
The linear environment is presented here as a parallel connection of
$LC$-oscillators, with the constraint $Z^{-1}(\omega) =
\sum_{\alpha,\pm} \left[iL_\alpha (\omega\pm\omega_\alpha+i0) \right]^{-1}$, 
where $\omega_\alpha=1/\sqrt{L_\alpha C_\alpha}$.

We obtained a Hamiltonian
description in terms of the phases $\theta_1$, $\theta_2$, $\phi$ and the
conjugate charges $en_1$, $-en_2$ and $q$, the latter being the total charge
passed through the voltage source relative to the equilibrium charge $C_{\rm
int}V$ on the plates of the capacitor $C_{\rm int}$.
At the relevant low frequencies the interaction with the bath reduces to $H_{\rm
int} = -Q_{\rm int}\delta V$, where
$\delta V \equiv (q-m)/C_{\rm int}$ is the fluctuating part of the transport
voltage. The dissipative Hamiltonian of the form
(\ref{eq:Dissipative_Hamiltonian}) provides for the proper high-frequency
regularization of the effective bosonic bath, with cut-off frequency $\omega_c =
(R C_{\rm int})^{-1}$, where $R$ is the real part of $Z(\omega)$.

\begin{figure}
\centerline{\hbox{\psfig{figure=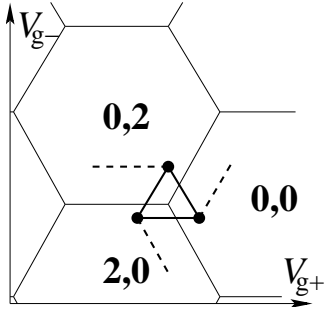,width=0.5\columnwidth}}}
\caption[]{\label{F:HoneyComb}FIG.\ref{F:HoneyComb}.
The honeycomb stability diagram of charge states. The solid dots and dashed 
lines denote the resonance peaks and lines, respectively}
\end{figure}

\noindent{\bf Qualitative analysis of the low-voltage resonances.}
In this section we provide qualitative analysis of the transport properties and
illustrate the discussion by the results of numerical simulation described
below.  We study resonances at transport voltages below the superconducting gap,
$eV < \Delta$ but assume that the voltage is high enough so that the features
related to the supercurrent through the system are not relevant.
The discussion and figures correspond to a positive bias $V>0$.
To understand the origin of possible resonances, let us first discuss the
stability diagram for the charge states (see Fig.~\ref{F:HoneyComb}),
neglecting the Josephson couplings.

In the unbiased case the stability conditions
define a honeycomb pattern in the gate-voltage plane.  Inside each
hexagon a certain charge state has the minimal energy.
At the vertices three charge states are degenerate. When a
transport voltage $V$ is applied these points grow into triangles, within which 
the system is unstable w.r.t. sequential tunneling of
Cooper pairs: $\ket{0,2,m}\to\ket{0,0,m}\to\ket{2,0,m}\to\ket{0,2,m+2}\dots$ [In
the experimentally relevant limit of Josephson couplings and temperatures below
the charging energy, in the vicinity of one vertex only three charge states 
$\ket{n_1,n_2}$ are
relevant: $\ket{2,0}$, $\ket{0,2}$ and $\ket{0,0}$.]
However, this gives a low current since the incoherent tunneling through the 
left and right junctions $\tilde E_{\rm J}$ is slow.

\begin{figure}
\centerline{\hbox{\psfig{figure=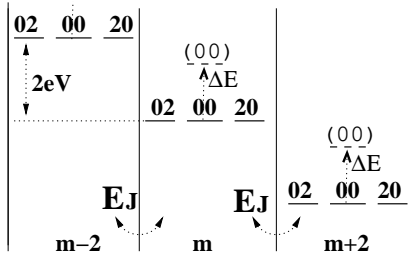,width=0.64\columnwidth}}}
\caption[]{\label{F:Charge_Res}FIG.\ref{F:Charge_Res}. Three charge levels at 
resonance. The {\sf (00)} 
and dotted arrows denote a passage along a resonant line}
\end{figure}

A much higher current can be achieved in resonant situations. One can expect
resonant points (peaks) and lines in the $V_{\rm g\pm}$-plane. At the peaks, 
defined by two constraints on $V_{\rm g\pm}$, three charge states are in
resonance. On the lines, only two charge states are degenerate.
The resonant conditions determine the positions of
possible peaks and lines. To evaluate the current at the resonant
peaks, we note that for typical parameters the bottleneck of
transport is associated with the incoherent transitions between triples of
resonant states; the rate of these transitions is given by the golden
rule and defines the current. However, the analysis of the shape of the
peaks/lines (the decay of current away from resonances) is more subtle. It may 
require the analysis of the Zeno regime and of the crossover to this
regime. Below we develop a suitable master-equation approach. We begin with a
qualitative discussion of the results.

Consider, for instance, the three-state resonance shown in
Fig.~\ref{F:Charge_Res}, which corresponds to
the upper vertex of the triangle in Fig.\ref{F:HoneyComb}. In this case
the Cooper pairs tunnel incoherently in the central junction only, and 
coherently through two other junctions.
The coherent couplings $\tilde E_{\rm J}$ exceed the rate of incoherent
transitions, which can be evaluated using the golden rule:
\begin{equation}
\label{eq:Gamma_r} \Gamma_{\rm r} \approx \frac{4\pi}{9} \frac{R}{R_{\rm Q}} 
\frac{E_{\rm J}^2}{2eV} \,,
\end{equation}
where $R_{\rm Q}\equiv h/(2e)^2$ and we assumed $T \ll 2eV$. This
rate defines the current magnitude at resonance, $2e \Gamma_{\rm
r}$.

Tuning the gates away from this resonance peak, one may still keep two levels
degenerate along a resonant line. For instance, one may lift the state 
$\ket{0,0,m}$ with respect to the degenerate $\ket{0,2,m}$ and $\ket{2,0,m}$ 
(see Fig.\ref{F:Charge_Res}; if $\ket{0,0}$ descends, the system may get 
Coulomb-blocked in this state).
In this configuration the transport involves a second-order coherent tunneling 
(cotunneling) $\ket{0,2,m}\to\ket{2,0,m}$ and incoherent relaxation 
$\ket{2,0,m}\to\ket{0,2,m+2}$.
To estimate the current, we evaluate the second-order
coherent coupling between $\ket{0,2,m}$ and $\ket{2,0,m}$ and find $\delta \sim 
\tilde E_{\rm J}^2/\Delta E$, where $\Delta E$ denotes the distance to the
$\ket{0,0}$-state (see Fig.\ref{F:Charge_Res}).
As discussed above on p.\pageref{Zeno-intro} the current remains to be 
$2e\Gamma_{\rm r}$ as long as $\delta>\Gamma_{\rm r}$.
However, for $\delta<\Gamma_{\rm r}$ the system is in the Zeno regime,
and the relaxation rate $\ket{2,0}\to\ket{0,2}$ defines
the current $\sim 2e\delta^2/\Gamma_{\rm r}$. Thus along the resonant line the 
current stays at the peak level and then drops fast. The deviation from the peak 
at which the current drops can be estimated from the condition 
$\delta\sim\Gamma_{\rm r}$; further behavior is governed by the Zeno physics. 
For the typical parameters~\cite{Bibow_Lafarge_Levy_PRL} (see below) this gives 
a very short line (it is also very narrow, cf.~below). This may explain why this 
resonant line was not detected.

If the threefold degeneracy in Fig.\ref{F:Charge_Res} is lifted in other ways 
(with two states still in resonance) the transport 
involves higher-order incoherent processes and the current is much 
weaker~\cite{Bibow_Thesis}. However, there exist other resonant peaks, which are 
located at two lower vertices of the triangle: one can say that in 
Fig.\ref{F:Charge_Res} the voltage drops at the central junction, but it can 
also drop at the left/right junction. The respective rate of incoherent 
Cooper-pair tunneling can be evaluated using Eq.~(\ref{eq:Gamma_r}) with the
substitution $E_{\rm J}\to \tilde E_{\rm J}$, i.e., the current at these peaks
is much lower. However, our analysis shows that the resonance lines
originating from these peaks are much longer (and wider, cf.~below) and may
reach the neighboring hexagons' vertices, as it was indeed found
experimentally. The reason is that at these peaks the incoherent rate is much
lower, the coherent coupling stronger, and it takes a longer distance away
from the peak for the coherent coupling to fall below the incoherent rate
(crossover to the Zeno regime).
Thus we find, in agreement with experiment, that only oblique
(but not horizontal) resonant lines should be visible and allows us to evaluate 
the shape of the resonances.

The widths of the resonant lines were evaluated in a similar way, with results 
in, at least, semi-quantitative agreement with experiment. We
remark that the width is not set by the condition of resonance as such (which
requires a charge-level splitting lower than the coupling and would
define very narrow lines~\cite{Bibow_Thesis}). In fact, during the separation of 
two resonant states the transport changes from coherent to incoherent. 
At this crossover point the incoherent rate is higher than the respective 
$\Gamma_{\rm r}$. Only at a longer distance from the line it drops below 
$\Gamma_{\rm r}$ and slows the transport. The respective width scales linearly 
with $V$, similar to the experiment~\cite{Bibow_Lafarge_Levy_PRL}.

So far we analyzed transport at voltages $V$ much higher than the
Josephson couplings and used the charge basis.  Now we focus on transport at 
lower voltages and find that due to the coherent Josephson
coupling of the charge states the triple resonance of
Fig.\ref{F:Charge_Res} appears only at voltages above a
certain threshold.

\begin{figure}
\centerline{\hbox{\psfig{figure=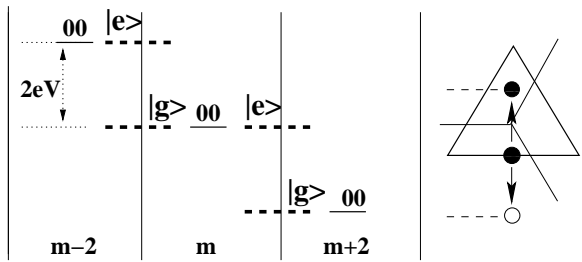,width=0.9\columnwidth}}}
\caption[]{\label{F:Eigenres}FIG.\ref{F:Eigenres}.
Resonances with the double-island's eigenstates.
The right panel shows the peaks' positions in the $V_{\rm g\pm}$-plane. A peak 
emerges at $V=E_{\rm J}/2e$ (the solid dot in the middle) and splits as the bias 
$V$ increases. The dashed lines show the cotunneling resonances}
\end{figure}

At lower $2eV\sim E_{\rm J}$
it is convenient to work in the eigenbasis of the double island.
Near the triangle in Fig.~\ref{F:HoneyComb} the difference $U$ in charging 
energies of the states $\ket{2,0,m}$ and $\ket{0,2,m+2}$ is small, and one finds 
the ground and excited eigenstates of
the double island,
\begin{eqnarray*}
\ket{g,m+2} = \cos\gamma\ket{2,0,m} &+&
\sin\gamma\ket{0,2,m+2}\,,\\
\ket{e,m} = -\sin\gamma\ket{2,0,m} &+&
\cos\gamma\ket{0,2,m+2}
\,,
\end{eqnarray*}
where $\tan 2\gamma = E_{\rm J}/U$.
 
The respective resonance configuration is shown in
Fig.\ref{F:Eigenres}.  Since the minimal energy splitting
between the ground and excited states is $E_{\rm J}$, the resonant conditions of
Fig.~\ref{F:Eigenres} require $2eV \ge E_{\rm J}$.  At $V
= E_{\rm J}/2e$ the peak is located at the lower side of the
triangle (see Fig.~\ref{F:Eigenres}). Above this threshold the equation
$E_{\rm e} - E_{\rm g} = E_{\rm J}/\sin2\gamma = 2eV$ ($0 < \gamma < \pi/2$) has
two solutions, and the peak splits:  The main peak with $\gamma > \pi/4$ enters 
the triangle, and the other, secondary peak ($\gamma < \pi/4$) leaves it. At 
strong bias $2eV \gg E_{\rm J}$ the main peak reaches the upper vertex of the 
triangle, while the secondary becomes very narrow and joins one of the oblique 
resonant lines. (Notice that the triangle itself slides and grows with the 
increase of $V$.)

Let us estimate the current magnitude at these resonances. The relaxation rate
$\ket{e,m}\to\ket{g,m+2}$ is given by Eq.(\ref{eq:Gamma_r}), and the matrix 
element between the states $\ket{e,m}$ and $\ket{g,m}$ to $\ket{0,0,m}$ due to 
$H_{\rm J}$ is $E_{\rm coupl} = (\tilde E_{\rm J} /2)\sin\gamma$,
of order $\tilde E_{\rm J}$ for the main resonance $\bullet$ and weaker, 
$\sim\tilde E_{\rm J} E_{\rm J}/(2eV)$, for the other one.

If $\tilde E_{\rm J} \gg (R/R_{\rm Q})\;E_{\rm J}$ the incoherent relaxation
inside the double-island is the bottleneck (the slowest stage) of the transport
for both peaks, $\Gamma_{\rm r} \ll E_{\rm coupl}$, i.e., the peak height is
$I_{\rm max}\approx 2e\Gamma_{\rm r}$.  However, the peaks' sizes are different
due to different $E_{\rm coupl}$ and can be found from the analysis similar as 
above.  The external peak $\circ$ is much narrower at $2eV \gg E_{\rm J}$.

\begin{figure}
\centerline{\hbox{\psfig{figure=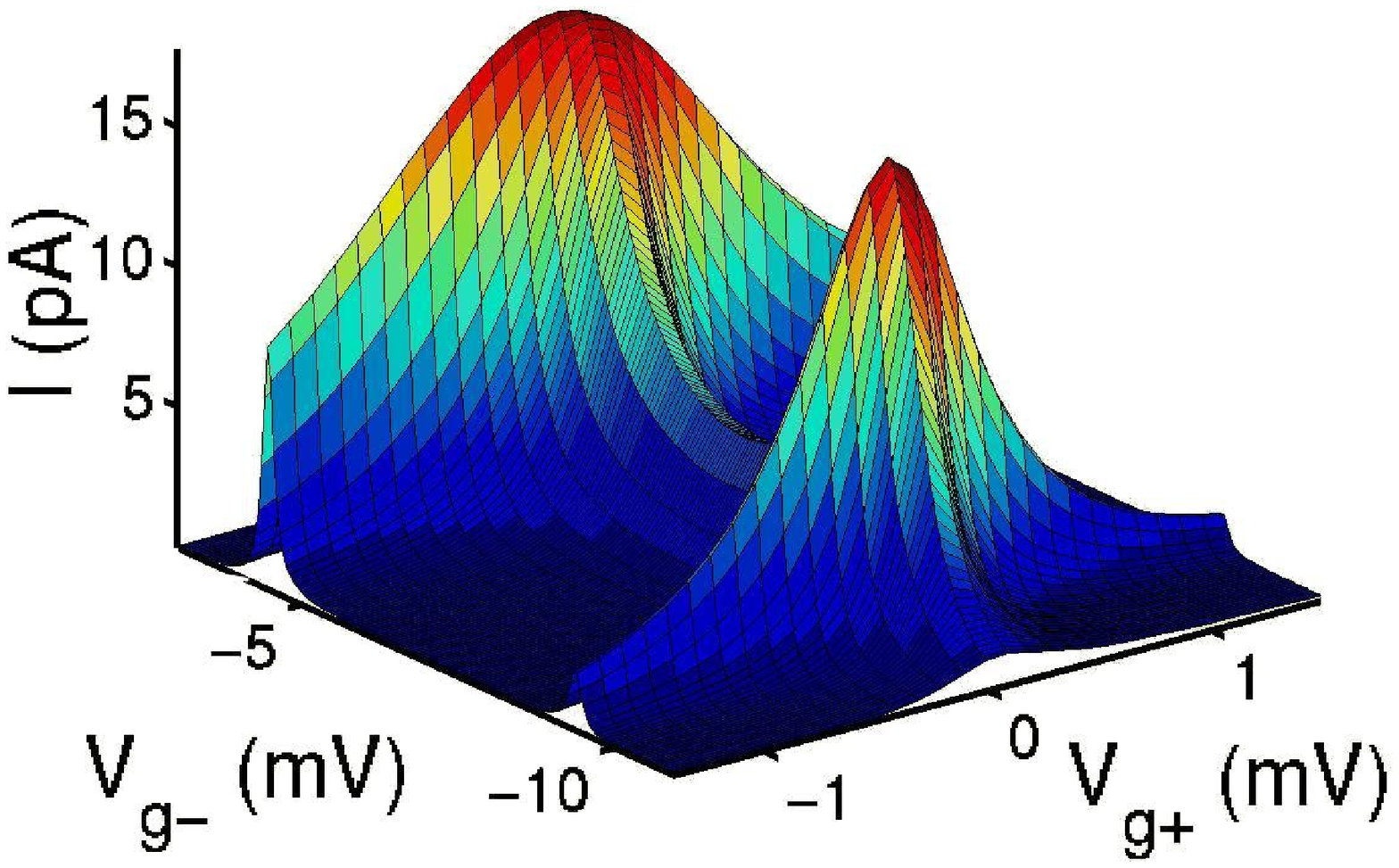,width=0.85\columnwidth}}}
\caption[]{\label{F:Results-23-50-full}FIG.\ref{F:Results-23-50-full}.
$I(V_{\rm g+},V_{\rm g-})$ for $V=23\mu$V and $T=50$mK}
\end{figure}

For $\tilde E_{\rm J} \ll (R/R_{\rm Q})\;E_{\rm J}$ one finds that
$\Gamma_{\rm r} \gg E_{\rm coupl}$
for the secondary peak, and also for the main
resonance at voltages $V$ close to $E_{\rm J}/2e$. The Zeno effect is
expected under these circumstances \cite{Harris_Stodolsky_Zeno}, the transport 
is slowed down, and transitions in the outer junctions define the current 
$I_{\rm max} \sim 2e E_{\rm coupl}^2/\Gamma_{\rm r}$.

\noindent{\bf Master equation and numerics.}
The dynamics reduce to propagation along the chain of eigenstates with 
decreasing energy and growing $m$. To evaluate the current we analyze the 
dynamics of the reduced density matrix $\hat\sigma$, retaining the indices
$n_1,n_2,m$ and tracing over the environment's degrees of freedom.
Using the real-time Keldysh diagrammatic technique
(cf.~Ref.~\cite{Schoeller_Schoen_PRB_94, Our_Book}), we find the
master equation
\begin{equation}
\label{eq:Dyson_Equation}
\frac{d}{dt}\hat\sigma(t)-L_0 \hat\sigma(t)=
\int\nolimits_{-\infty}^t dt'\; \Sigma(t-t')\;\hat\sigma(t')
\,,
\end{equation}
with the bare Liouville operator $L_0 \equiv i \left[ \cdot,
H_0 \right]$, $H_0=H_{\rm C} + H_{\rm J}$. In the first (Born) approximation we 
obtain
\begin{equation}
\Sigma(t) = \alpha'(t) L_{\rm int} e^{L_0 t} L_{\rm int} -
            i\alpha''(t) L_{\rm int} e^{L_0 t} M_{\rm int}
            \ ,
\label{eq:Sigma_First_Order}
\end{equation}
where $\alpha(t)=\alpha'(t)+i\alpha''(t)$ is given by
\begin{equation}
\alpha(t) \equiv (2e)^2\langle \delta V(t) \delta V(0) \rangle =\!
\int\! \frac{d\omega}{\pi}
\frac{J(\omega) e^{-i\omega t}}{1-e^{-\hbar\omega/k_{\rm B}T}}
\,,
\end{equation}
the low-frequency spectral density $J(\omega) = 2\pi \omega\,R/R_{\rm Q}$, and
$L_{\rm int} \equiv i \left[ \cdot , Q_{\rm int}/2e \right]$, $M_{\rm int} 
\equiv i \left[ \cdot , Q_{\rm int}/2e
\right]_+$.
The last term in Eq.~(\ref{eq:Sigma_First_Order}) violates the translational 
symmetry $m\to m+2$. The invariance is restored after a regularization, due to 
the counterterm $Q_{\rm int}^2/2C_{\rm int}$ in 
Eq.~(\ref{eq:Dissipative_Hamiltonian})~(cf.~Ref.~\cite{Caldeira_Leggett_83}).

We label the entries of the self-energy matrix $\Sigma$ by four triples
$\nu^\mp$ and  $\nu'^\mp$, where
e.g.  $\nu^-=(n_1^-,n_2^-,m^-)$.  Here the sign $\mp$ refers to a Keldysh 
branch; the (un)primed indices refer to the time $t'$ (resp. $t$).  Most of 
these indices vary over finite ranges.  Indeed, only the lowest charge states 
$n_1$, $n_2$ participate in the low-frequency dynamics, and strongly 
off-diagonal entries, with large
$m^-\!\!-m^+$ and $m'^-\!\!-m'^+$, are suppressed.
The regularized self-energy is translationally invariant and does not depend on 
the sum $m^-\!\!+m^++m'^-\!\!+m'^+$. The Fourier transform w.r.t. 
$(m^-\!\!+m^+-m'^-\!\!-m'^+)/2$ gives a finite
matrix for each value of $k$.

We use the Laplace-transformed master equation: $s\hat\sigma(k,s) - 
\hat\sigma_0(k) = \Pi(s,k)\hat\sigma(k,s)$ to find the current
$I = s^2\langle m(s)\rangle|_{s\to 0}$,
where $\langle m(s)\rangle=
i\partial_k{\rm Tr}\,\hat\sigma(k\!\!=\!\!0,s)
=i{\rm Tr}\,(s-\Pi)^{-1}\partial_k\Pi\;(s-\Pi)^{-1}\,\hat\sigma_0\,
|_{k\to 0}$. Here $\Pi(k,s) \equiv L_0(k) + \Sigma(k,s)$ and $\hat\sigma_0$ is 
the initial condition.
The numerical analysis can be simplified by taking the needed derivatives 
analytically and working in the eigenbasis of $H_0$. We ascribe a counting index 
$\tilde m$ to {\it eigenstates} (rather than charge states) and organize them 
into zones with fixed values of $\tilde m$~\cite{Maassen_Schoen_Geerligs_PRL_91,
Maassen_Odintsov_Bobbert_Schoen_ZPhysB_91}. The eigenstates of the total 
Hamiltonian~(\ref{eq:ham})
have only a finite $m$-spread, and one can use $\tilde m$ to evaluate the dc 
current.

In our analysis we used the following parameters: $C_{\rm J} =
0.8$\,fF, $C_{\rm g} = 8$\,aF, $\tilde E_{\rm J} = 25$\,mK, $E_{\rm J} = 
0.5$\,K, $R = 50\,\Omega$~\cite{Bibow_Thesis}. Fig.\ref{F:Results-23-50-full} 
shows the shape of two peaks and resonance lines for the bias $V$ just above the 
threshold $E_{\rm J} /2e \approx 21.5 {\rm \mu V}$, in agreement with estimates 
above.

\noindent{\bf Discussion.}  In our calculation we neglected the influence of the
$1/f$ noise due to background-charge fluctuations.  This very-low-frequency
noise dominates the pure dephasing (that leads to energy fluctuations without
transitions; its Ohmic part is included in our numerical analysis).  Our
estimates show that these effects should not change the results substantially.

In conclusion, using the methods that allow to cover the (Zeno) dynamics of
coherent systems under strong dissipation, we analyzed the Cooper-pair transport
through a double-island structure. We find separate peaks and resonant lines, 
whose visibility and shapes match the experimental observations. We further
predict a double-peak structure near a threshold transport voltage, observation
of which would be a probe of coherent properties of the double-island qubit.

We are grateful to E.~Bibow, P.~Lafarge, and L.~L\'evy for providing their
results and for numerous discussions.  This work is part of a research network
of the Landesstiftung BW.  Y.M.  was supported by the Humboldt foundation, the
BMBF, and the ZIP programme of the German government.

\end{document}